\documentclass[prl,twocolumn,aps,showpacs,preprintnumbers,amsmath,amssymb, superscriptaddress]{revtex4-1}
\pdfoutput=1

\usepackage{bm}
\usepackage{graphicx}
\usepackage{color}

\begin{document}

\title{Observation of Dirac-like band dispersion in LaAgSb$_2$}

\author{X. Shi}
\affiliation{Beijing National Laboratory for Condensed Matter Physics, and Institute of Physics, Chinese Academy of Sciences, Beijing 100190, China}

\author{P. Richard}
\email{p.richard@iphy.ac.cn}
\affiliation{Beijing National Laboratory for Condensed Matter Physics, and Institute of Physics, Chinese Academy of Sciences, Beijing 100190, China}
\affiliation{Collaborative Innovation Center of Quantum Matter, Beijing, China}

\author{Kefeng Wang}\thanks{Current address: CNAM, Department of Physics, University of Maryland, College Park, Maryland 20742, USA}
\affiliation{CMPMSD, Brookhaven National Laboratory, Upton, NY 11973 USA}

\author{M. Liu}
\affiliation{Beijing National Laboratory for Condensed Matter Physics, and Institute of Physics, Chinese Academy of Sciences, Beijing 100190, China}

\author{C. E. Matt}
\affiliation{Swiss Light Source, Paul Scherrer Institut, CH-5232 Villigen PSI, Switzerland}
\affiliation{Laboratory for Solid State Physics, ETH Z\"{u}rich, CH-8093 Z\"{u}rich, Switzerland}

\author{N. Xu}
\affiliation{Swiss Light Source, Paul Scherrer Institut, CH-5232 Villigen PSI, Switzerland}
\affiliation{Institute of Condensed Matter Physics, \'Ecole Polytechnique F\'ed\'erale de Lausanne, CH-1015 Lausanne, Switzerland}

\author{R. S. Dhaka}
\affiliation{Department of Physics, Indian Institute of Technology Delhi, Hauz Khas, New Delhi-110016, India}
\affiliation{Swiss Light Source, Paul Scherrer Institut, CH-5232 Villigen PSI, Switzerland}
\affiliation{Institute of Condensed Matter Physics, \'Ecole Polytechnique F\'ed\'erale de Lausanne, CH-1015 Lausanne, Switzerland}

\author{Z. Ristic}
\affiliation{Belgrade University, VINCA Institute of Nuclear Sciences, Atomic Physics Laboratory, POB 522, 11001 Belgrade, Serbia}
\affiliation{Swiss Light Source, Paul Scherrer Institut, CH-5232 Villigen PSI, Switzerland}
\affiliation{Institute of Condensed Matter Physics, \'Ecole Polytechnique F\'ed\'erale de Lausanne, CH-1015 Lausanne, Switzerland}

\author{T. Qian}
\affiliation{Beijing National Laboratory for Condensed Matter Physics, and Institute of Physics, Chinese Academy of Sciences, Beijing 100190, China}

\author{Y.-F. Yang}
\affiliation{Beijing National Laboratory for Condensed Matter Physics, and Institute of Physics, Chinese Academy of Sciences, Beijing 100190, China}
\affiliation{Collaborative Innovation Center of Quantum Matter, Beijing, China}

\author{C. Petrovic}
\affiliation{CMPMSD, Brookhaven National Laboratory, Upton, NY 11973 USA}

\author{M. Shi}
\affiliation{Swiss Light Source, Paul Scherrer Institut, CH-5232 Villigen PSI, Switzerland}

\author{H. Ding}\email{dingh@iphy.ac.cn}
\affiliation{Beijing National Laboratory for Condensed Matter Physics, and Institute of Physics, Chinese Academy of Sciences, Beijing 100190, China}
\affiliation{Collaborative Innovation Center of Quantum Matter, Beijing, China}

\date{\today}

\begin{abstract}
{We present a combined angle-resolved photoemission spectroscopy (ARPES) and first-principles calculations study of  the electronic structure of LaAgSb$_2$ in the entire first Brillouin zone. We observe a Dirac-cone-like structure in the vicinity of the Fermi level formed by the crossing of two linear energy bands, as well as the nested segments of Fermi surface pocket emerging from the cone. Our ARPES results show the close relationship of the Dirac cone to the charge-density-wave ordering, providing consistent explanations for exotic behaviors in this material.
}

\end{abstract}



\maketitle


Relativistic Dirac fermions occur in condensed matter physics when the electronic dispersion takes a cone-like shape called Dirac cone, which leads to various quantum phenomena. Dirac cones appear naturally as protected conducting surface states in topological insulators (TIs) \cite{Hasan_TI_RMP} and in graphene \cite{Castro_graphene_RMP}, but they have also been predicted and observed in other materials such as BaFe$_2$As$_2$, which is the parent compound of one family of Fe-based superconductors \cite{Ran_PRB79,Richard_DiracCone,Harrison_PRB80}. Recently, possible Dirac fermions were also reported in layered rare-earth silver antimonide LaAgSb$_2$ \cite{Wang_LaAgSb2}, a material showing interesting transport and magnetic properties \cite{Myers_PRB60,Myers_JMMM205}. In particular, LaAgSb$_2$ has a large linear magnetoresistance (MR) and a positive Hall resistivity, but its Seebeck coefficient is negative in zero magnetic field. Based on electronic structure calculations indicating that the linear $p$ bands from different Sb atoms cross each other to form a cone-like structure, it was proposed that these unusual properties could be attributed to the quantum limit of possible Dirac fermions \cite{Wang_LaAgSb2}. Unfortunately, a direct observation of Dirac cones electronic structure in LaAgSb$_2$ is still missing, despite a previous ARPES study focused around the Brillouin zone (BZ) center \cite{Arakane_LaAgSb2_ARPES}.

In this paper we report a comprehensive angle-resolved photoemission spectroscopy (ARPES) investigation and first-principles calculations of the electronic band structure of LaAgSb$_2$. The results clearly reveal the existence of linear energy bands and a Dirac-cone-like structure between the center and the boundary of the first BZ. Interestingly, a charge-density-wave (CDW) is observed on the bands emerging from the Dirac cone. These features provide natural explanations for various exotic behaviors previously reported in this material, including the existence of CDW ordering.



Single crystals of LaAgSb$_2$ were grown by the high-temperature self-flux method \cite{Wang_LaAgSb2}. ARPES measurements were performed at the Surface and Interface Spectroscopy (SIS) beamline of Swiss Light Source using a VG-Scienta R4000 electron analyzer. The energy and momentum resolutions have been set to 10$\sim$15 meV and 0.2$^{\circ}$, respectively. Fresh surfaces for the ARPES measurements were obtained by cleaving the samples \emph{in situ} in a working vacuum better than 8 $\times$ 10$^{-11}$ Torr. All data shown in this paper were recorded at 16 K. First-principles calculations using the full-potential linearized augmented plane-wave method as implemented in WIEN2K \cite{WIEN2K} were performed to investigate the electronic properties of LaAgSb$_2$. Lattice parameters and atomic positions obtained experimentally are used for the numerical calculations \cite{Sologub_JSSC}. We used the GGA-PBE exchange correlation energy and a 50000 $k$-points mesh for the first BZ. We used 2.50 a.u. for the muffin-tin radii of La, Ag and Sb. The spin-orbit coupling (SOC) is also included in the nonmagnetic band structure calculations.


\begin{figure*}
    \includegraphics[width=6in]{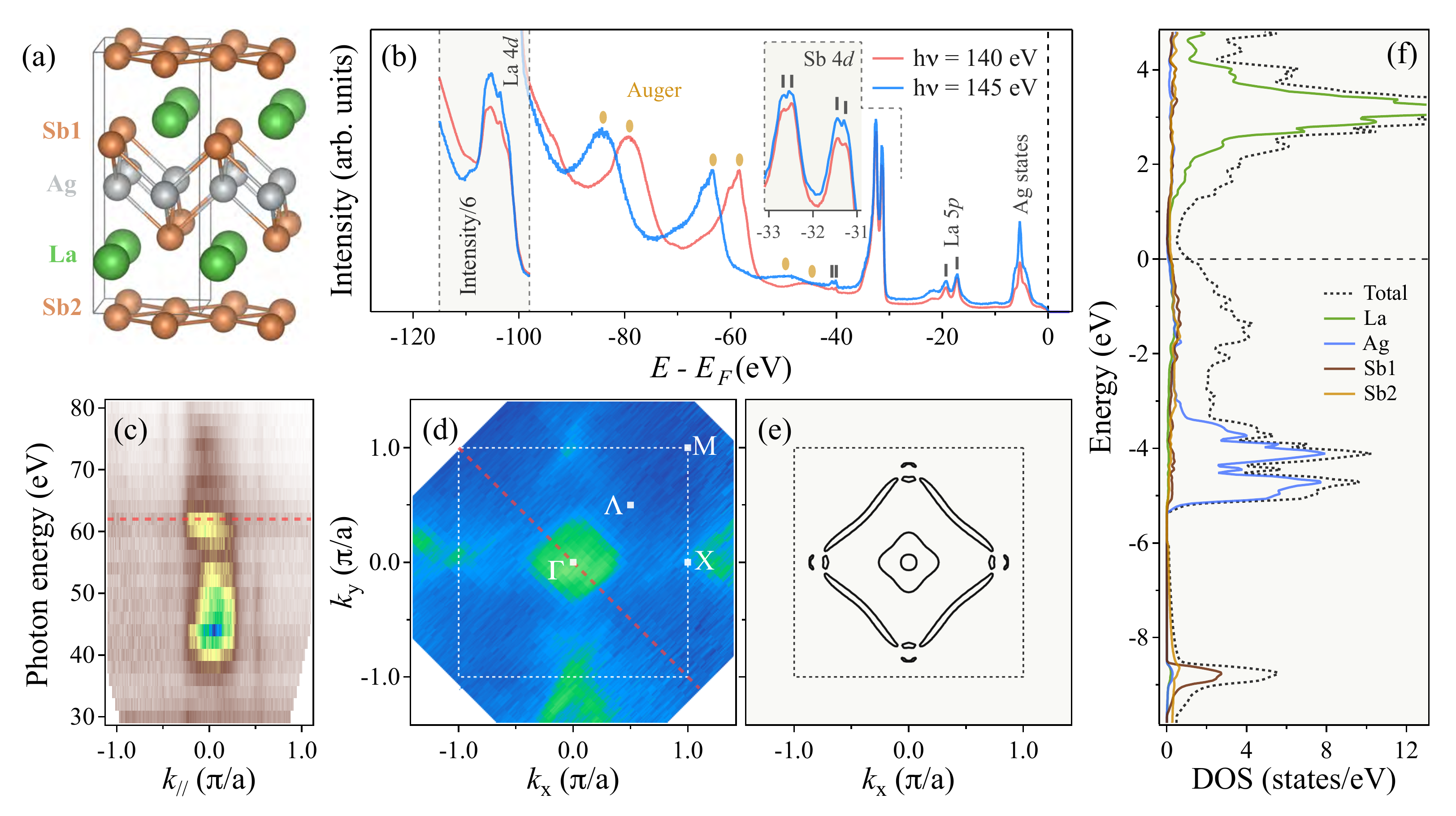}
    \caption{\label{Fig1}(Color online) (a) Crystal structure of LaAgSb$_2$. (b) Core level spectra recorded with 140 eV (red) and 145 eV (blue) photons. The insets show the La 4$d$ peaks around -104 eV and Sb 4$d$ peaks around -32 eV. (c) Photon energy dependence of the ARPES intensity plot along the $\Gamma$-M symmetry line. The red dashed line indicates the photon energy of 62 eV chosen to record the band structure and FS mappings. (d) FS intensity map recorded with 62 eV photons and integrated within a 20 meV energy window with respect to $E_F$. (e) Calculated FSs with SOC taking into account. (f) Calculated total DOS (dashed black line) and partial DOS projected onto the La, Ag, Sb1 and Sb2 states in LaAgSb$_2$.
}
\end{figure*}

As illustrated in Fig. \ref{Fig1}(a), LaAgSb$_2$ crystallizes in a tetragonal crystal structure (P4/nmm) consisting of La ions intercalated between Sb and AgSb layers. The elemental composition of our LaAgSb$_2$ samples is confirmed by the core level spectra displayed in Fig. \ref{Fig1}(b). Two photon energies (140 eV and 145 eV) were chosen during the measurements to distinguish direct photoemission peaks from peaks induced by light generated from the Auger process, for which the kinetic energy ($E_k$) of the photoelectron is constant with respect to the incident photon energy. Starting from the low binding energies (right side in Fig. \ref{Fig1}(b)), we first observe a large photoemission intensity around -5 eV, which we attribute mainly to Ag $5d$ states according to previous calculations \cite{Inada_PMB82} and to the calculated density-of-states (DOS) plotted in Fig. \ref{Fig1} (f). Then come the La $5p$ states, with peaks located at binding energies of 19.3 eV and 17.2 eV. The strongest photoemission intensity is found for the Sb $4d$ electrons, with peaks at $\sim$32.6 eV (Sb 4$d_{3/2}$) and $\sim$31.4 eV (Sb 4$d_{5/2}$). As illustrated in the right inset of Fig. \ref{Fig1}(b), each Sb $4d$ peak is split by about 0.2 eV, most likely due to the existence of two inequivalent Sb sites in the crystal.

Besides the photoemission peaks that we can ascribe unambiguously, the spectra have strong contributions from Auger peaks associated with kinetic energies $E_k$ of 57 eV, 77.8 eV and 91.4 eV. We also observe a weak peak around 42 eV, whose origin may be related to a satellite photoemission peak of the Sb $4d$ levels. Interestingly, the photon energy dependence from 30 eV to 80 eV of the ARPES intensity at the Fermi level ($E_F$) along the $\Gamma$-M high-symmetry line, displayed in Fig. \ref{Fig1}(c), shows a resonance at about the same energy (42 eV). The same plot is also a measure of the quasi-two-dimensionality of the compound. Indeed, within the sudden approximation and the nearly-free electron model for the photoemission process, the photon energy is directly related to the out-of-plane dispersion \cite{Hufner_Photoemission,Richard_JPCM27}. In agreement with the nearly two-dimensional nature of the crystal structure \cite{Myers_PRB60}, we observe only very small dispersion along $k_z$, except for the most inner pocket centered at $\Gamma$, which is more three-dimensional. Because of the quasi-two-dimensionality of the compound, we cannot identify $k_z$ unambiguously.

\begin{figure*}
  \begin{minipage}{0.7\textwidth} 
    \includegraphics[width=4.5 in]{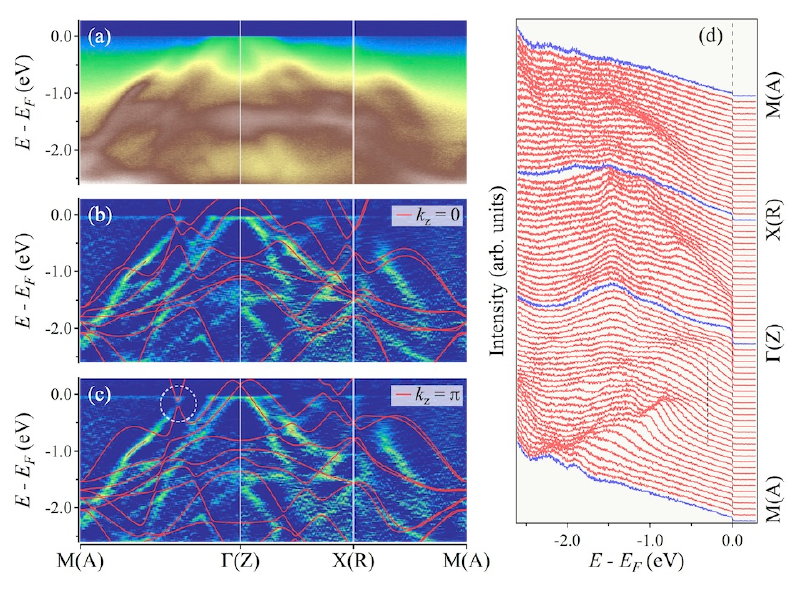}
  \end{minipage}%
  \begin{minipage}{0.3\textwidth}
    \caption{\label{Fig2}(Color online) (a) ARPES intensity plot of LaAgSb$_2$ along the M(A)-$\Gamma$(Z)-X(R)-M(A) high symmetry lines recorded with 62 eV photons. 
    (b) Corresponding two-dimensional curvature plot. The calculated band dispersions (red lines) in the $k_z~=~0$ plane are plotted for comparison. (c)~Same as (b) but with the calculated band dispersions in the $k_z~=~\pi$ plane appended. The circle in (c) indicate the position of the Dirac-cone-like structure.
    (d) EDC plot corresponding to the data in panel (a). The segment of dashed line 300 meV below $E_F$ is a guide to show a strong reduction of the spectral intensity.
}
  \end{minipage}
\end{figure*}

We display in Fig. \ref{Fig1}(d) the Fermi surface (FS) intensity map recorded with 62 eV photons. In agreement with a previous ARPES study \cite{Arakane_LaAgSb2_ARPES}, strong photoemission intensity is found at the $\Gamma$ point. As illustrated in Fig. \ref{Fig2}, this intensity comes from two hole pockets. We observe an additional diamond-like pocket with its corners at the X point that has not been reported by ARPES in this system.
This FS differs but is similar to that of isostructural SrMnBi$_2$, which consists of four separated needlelike pockets along the $\Gamma$-M direction \cite{Park_SrMnBi2, LLJia_SrMnBi2, XJZhou_SrMnBi2, GeunsikLee_Calculations_AMnBi2}, and with that of  YbMnBi$_2$ \cite{Borisenko_arxiv}, which as almost the same crystal structure. Interestingly, similar FS topology has also been reported recently on materials exhibiting exotic electronic states attributed to line-node Dirac cone surface state in ZrSiS \cite{Schoop_arxiv} and to a weak topological insulator in ZrSnTe \cite{Shancai_arxiv}. Our results on LaAgSb$_2$ are consistent with a previous de Haas-van Alphen study on this material \cite{Myers_PRB60}. Our first-principles calculation results plotted with black lines in Fig. \ref{Fig1}(e) capture the FS topology well except for some tiny features around X. According to the calculations, the diamond-like pocket mainly consists of two parts very close and nearly parallel to each other, and originating from Sb $5p_{x}$ and $5p_{y}$ states. This leads to good FS nesting conditions, as further discussed below.

To systematically investigate the electronic structure of LaAgSb$_2$, we recorded high-resolution and high-statistic photoemission cuts with 62 eV photons along the high-symmetry lines M(A)-$\Gamma$(Z)-X(R)-M(A) in the first BZ, as displayed in Fig. \ref{Fig2}(a). We display the corresponding  two-dimensional curvature \cite{PZhang_Curvature} plot in Fig. \ref{Fig2}(b)-(c) and the energy distribution curves (EDCs) in Fig. \ref{Fig2}(d) to highlight the band dispersions. 
We note that the spectral feature are not extremely sharp due to the intrinsic difficulty to obtain very a flat cleaved surface for this soft compound. Moreover, the spectral intensity at low energy is reduced by a CDW, as discussed below.
The calculated bands in both $k_z = 0$ and $k_z = \pi$ planes are overlapped on the curvature plots for comparison. We found that the measured electronic structure at low-energy matches better to the calculated one at $k_z = \pi$, but the exact $k_z$ position is not certain, as mentioned above. Without any renormalization, the overall calculated bands capture most of the experimental features, suggesting that the electronic correlations are weak in LaAgSb$_2$.

\begin{figure}
\begin{center}
\includegraphics[width=3.5in]{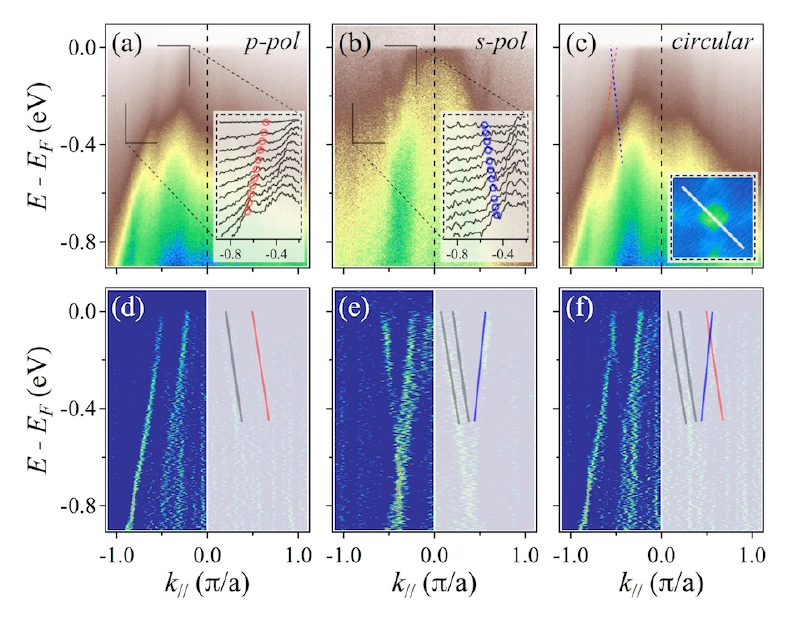}
\end{center}
\caption{\label{Fig3}(Color online) (a)-(c) ARPES intensity plot along the $\Gamma$-M direction, as indicated in the inset of panel (c), recorded with 62 eV $p$-polarized, $s$-polarized and circularly-polarized photons, respectively. The insets in (a) and (b) show the MDCs in the marked area. The red and blue open circles indicate the MDC peaks and show the linear band dispersions. These two bands with different orbital characters cross and form a Dirac-cone-like band structure. (d)-(f) Intensity plots of MDC curvature corresponding to the data presented in (a)-(c), respectively. The extracted band dispersions are appended on the plots.
}
\end{figure}

We now focus on the Dirac-cone-like band structure that has been proposed for this material \cite{Wang_LaAgSb2}. As highlighted by a circle in Fig. \ref{Fig2}(a), our own calculations indicate two nearly linear bands with Sb $5p_{x}$ and $5p_{y}$ characters intersecting at ($\pm$0.39, $\pm$0.39)$\pi/a$, which is slightly away from the mid-point between $\Gamma$ and M ($\Lambda$ point in Fig. \ref{Fig1}(d)). In Fig. \ref{Fig3} we show a zoom on the ARPES intensity plot at low binding energy along the $\Gamma$-M direction, as indicated in the inset of \ref{Fig3}(c).
Taking advantage of the photoemission matrix element effect \cite{XPWang_OrbitalCharacters}, we tune the light to linear polarization to identify the orbital symmetry of the bands. As shown in Figs. \ref{Fig3}(a) and (b), the $p$ and $s$ configurations highlight the intensity of different bands. The momentum distribution curves (MDCs) displayed in the insets and the momentum curvature plots (Figs. \ref{Fig3}(d) - \ref{Fig3}(f)) show these bands more clearly. In each spectrum we observe linear band dispersions, as marked by red and blue open circles in the MDC plots. In Fig. \ref{Fig3}(c), we display a spectrum recorded using circular polarized light, which combines features observed by the $s$ and $p$ configurations. These two highlighted bands cross each other and form a Dirac-cone-like band structure, as expected from the calculations.



Experimentally, we find that the Dirac point locates $\sim$ 100 meV below $E_F$, while it is found $\sim$ 160 meV below $E_F$ in the calculations. The calculations show a $\sim$ 60 meV gap at the Dirac point, whereas no gap is detected by ARPES within the experimental resolution. We find the Dirac point at $k_{\parallel}$ = ($\pm$0.38, $\pm$0.38)$\pi/a$ in the momentum space. 
As with SrMnBi$_2$ \cite{Park_SrMnBi2, LLJia_SrMnBi2, XJZhou_SrMnBi2, GeunsikLee_Calculations_AMnBi2} and BaFe$_2$As$_2$ \cite{Richard_DiracCone}, and in contrast to graphene \cite{TOhta_Graphene}, this location of the Dirac cone is not a high-symmetry point, thus relaxing the requirements of isotropy for the Dirac cone. The cone observed here is strongly anisotropic. The band with even symmetry (selected by the $p$ configuration) has a Fermi velocity $v_F$ $\approx$ 3.75 eV \r{A} (5.7 $\times$ 10$^{5}$ m/s), while the other one has $v_F$ $\approx$ 5.24 eV \r{A} (8 $\times$ 10$^{5}$ m/s). The calculations give $v_F$ values of 10.2 eV \r{A} and 8.4 eV \r{A}, respectively. Although we cannot extract $v_F$ precisely in the perpendicular direction (roughly along $\Lambda$-X), we can estimate it from our knowledge of the energy position of the Dirac cone and by assuming that the band emerges at X. We found that $v_F$ along $\Lambda$-X is about 20 times smaller than in the other directions. In SrMnBi$_2$ the asymmetry of the Dirac cone was attributed to the spin-orbital coupling \cite{XJZhou_SrMnBi2}. The $v_F$ of LaAgSb$_2$ is smaller compared to that of SrMnBi$_2$ (1.1 $\times$ 10$^{6}$ m/s along the $\Gamma$-M direction), in agreement with the analysis of the critical field in the MR measurements \cite{Wang_LaAgSb2, KFWang_SrMnBi2}. 

\begin{figure}
\begin{center}
\includegraphics[width=3.5in]{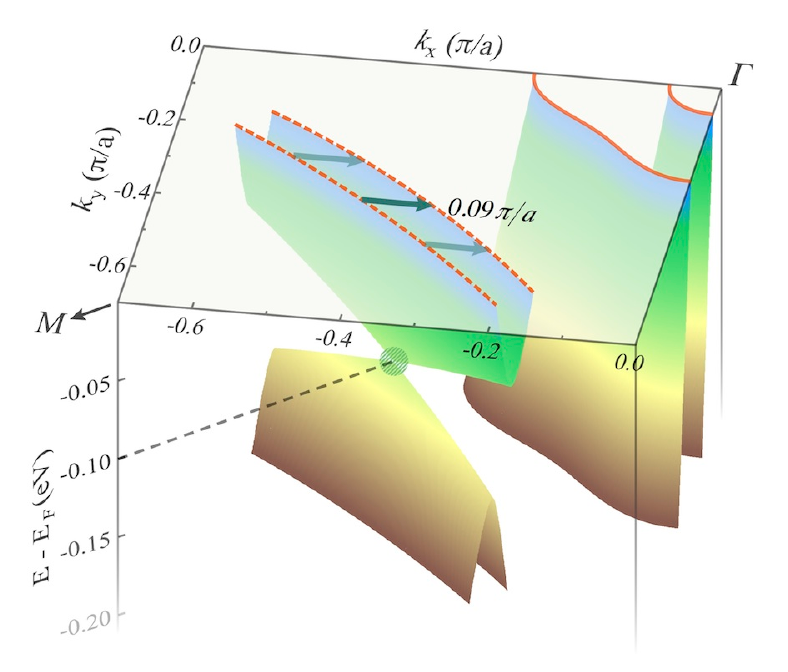}
\end{center}
\caption{\label{Fig4}(Color online) Schematic plot of the three-dimensional band dispersion in a part of the first BZ. The segments of FS pocket associated to the Dirac-cone-like structure are well nested with a small wave vector as indicated by the cyan arrows. 
}
\end{figure}

Our main results on the electronic structure in LaAgSb$_2$ are summarized in Fig. \ref{Fig4}. Along the $\Gamma$-M high-symmetry line, four bands cross $E_F$. The three-dimensional representation of the band dispersions is illustrated based on the experimental parameters. The Dirac-cone-like dispersion is formed by the intersection of the two outer linear bands associated with the diamond-like FS shown in Fig. \ref{Fig1}(d).
As mentioned above, the parallel portions of this FS are well nested, offering a straightforward explanation for the CDW ordering at $\sim207$ K suggested by a previous X-ray scattering study \cite{Song_LaAgSb2_CDW}. This temperature coincides with anomalies at 207 K observed in the resistivity and magnetic susceptibility \cite{Myers_JMMM205}. 
As sketched in Fig. \ref{Fig4}, the small wave vector $\vec{q}$ connecting the nested portions of the FS is estimated from our experimental data to be (0.09$\pm$0.04, 0)$\pi/a$, which is comparable to the modulation wave vector (0.052, 0)$\pi/a$ in the X-ray scattering measurements. Interestingly, the spectral weight of the bands emerging from the Dirac cone and forming the elongated pockets is quite reduced over an energy range of about 300 meV, as particularly well illustrated by the EDC plot in Fig. \ref{Fig2}(c), providing further indication that these bands are indeed responsible for the CDW ordering. We note that in a previous MR measurement \cite{Wang_LaAgSb2}, the linear MR, which is suggested to relate to the Dirac-like structure, disappears above the CDW transition temperature, which may also give some hint to the relationship between the Dirac fermions and the CDW ordering in LaAgSb$_2$.

In summary, we report the direct observation of the linear Sb $5p_{x,y}$ energy bands, and the Dirac-cone-like structure located along the $\Gamma$-M direction, which originates from the crossing of these bands. Interestingly, the FS pockets emerging from the Dirac cone are well nested with a small wave vector. The unique electronic band structure leads to the existence of Dirac Fermions and CDW ordering in LaAgSb$_2$, and further provides natural explanations for various novel transport and magnetic properties.


We acknowledge S.-F. Wu and J.-Z. Ma for useful discussions. This work was supported by grants from MOST (2011CBA001000, 2011CBA00102, 2012CB821403, 2013CB921703 and 2015CB921301) and NSFC (11004232, 11034011/A0402 and 11274362) from China. Work at BNL was supported by the Materials Sciences and Engineering Division, Office of Basic Energy Sciences, U.S. DOE under Contract No. DE-SC00112704. This work was supported by the Sino-Swiss Science and Technology Cooperation (No. IZLCZ2138954), and the Swiss National Science Foundation (No.200021-137783).

\bibliography{bibLAS}
\end{document}